\documentclass[conference]{IEEEtran}
\IEEEoverridecommandlockouts
\let\oldbibliography\thebibliography
\renewcommand{\thebibliography}[1]{%
  \oldbibliography{#1}%
  \setlength{\itemsep}{0pt}}   

\usepackage{tikz}
\usetikzlibrary{
    arrows.meta,
    positioning,
    shapes.geometric,
    calc
}
\usepackage{soul}
\usepackage{xcolor}
\usepackage{amsmath,amssymb}   

\usepackage{multirow}
\usepackage{cite}
\usepackage{amsmath,amssymb,amsfonts}
\usepackage{algorithmic}
\usepackage{algorithm}
\usepackage{graphicx}
\usepackage{amsmath}
\usepackage{textcomp}
\usepackage{xcolor}
\usepackage{url}
\usepackage{wrapfig}
\usepackage{balance}
\usepackage{subcaption}
\usepackage[font=footnotesize,labelfont=bf]{caption}
\usepackage{booktabs,tabularx,graphicx}

\begin{document}
\title{
{Multi-Library Feature Engineering and Statistically Validated Classification for Power Swing Detection and Stability Assessment} }
\author{
{\tt\small}
Pallav Kumar Bera\\
Western Kentucky University, KY, USA \\
{\tt\small pallav.bera@wku.edu\vspace*{-0.1 cm}}
\and
Samita Rani Pani\\
KIIT DU, Odisha, India\\
{\tt\small samita.panifel@kiit.ac.in\vspace*{-0.1 cm}}
\and
Ranjeeta Patel\\
KIIT DU, Odisha, India\\
{\tt\small ranjeeta.patelfel@kiit.ac.in\vspace*{-0.1 cm}}

\thanks{ © 2026 IEEE.  Personal use of this material is permitted.  Permission from IEEE must be obtained for all other uses, in any current or future media, including reprinting/republishing this material for advertising or promotional purposes, creating new collective works, for resale or redistribution to servers or lists, or reuse of any copyrighted component of this work in other works.}}

\maketitle

\begin{abstract}
Faults occurring during power swings, and the evolution of a swing into an unstable condition, cause distance-relay maloperation and cascaded outages. This paper identifies faults, faults during power swings, and power swings, and distinguishes stable from unstable swings in a 9-bus WSCC system modeled in PSCAD/EMTDC. Three time-series feature libraries (tsfresh, TSFEL, and tsfeatures) are applied to the 3-phase relay voltages and currents, giving 3286 statistical, temporal, spectral, and fractal descriptors per event window. A two-step feature selection (Boruta all-relevant filtering followed by RFECV- and SHAP-guided minimal-optimal pruning) reduces the feature set to 18 compact, non-redundant features. 14 classifiers spanning linear, kernel/instance-based, probabilistic, neural, and tree-ensemble/boosting families are then benchmarked under repeated stratified cross-validation, with model differences assessed by a Friedman/Nemenyi omnibus-and-post-hoc test and a Nadeau--Bengio corrected paired $t$-test. Rotation Forest (ROF) identifies the three event types with 99.98\%  and SVM-RBF distinguishes stable from unstable swings with 97.97\% balanced accuracy. The results show that automated, library-driven feature extraction combined with rigorous statistical model comparison achieves near-perfect discrimination while retaining feature-level interpretability through SHAP importance.
\end{abstract}


\section{Introduction}
\label{sec:introduction}

Distance relays remain the primary protection for high-voltage transmission lines, but their impedance-based operating logic is vulnerable to two related disturbances. During a power swing, the apparent impedance trajectory can encroach the relay's protective zones and cause an undesired trip even though no fault exists. Also, if an actual fault occurs while a swing is already present, a conservative power-swing-blocking (PSB) scheme can delay or prevent the trip. This discrimination is hardest for symmetrical (3-phase) swings, since a balanced swing produces voltage and current waveforms that resemble that of a 3-phase fault \cite{mohammadi2022psb}. Asymmetrical swings, by contrast, are comparatively easier to unblock, since their zero- and negative-sequence components reveal the imbalance; however, they are becoming more common as single-pole tripping and reclosing on heavily loaded, long lines briefly unbalance the system during the open-pole interval \cite{mohammadi2022psb, arumuga2022distance}. Swings can be stable or unstable: in a stable swing the rotor-angle oscillation is underdamped and the system settles into a new equilibrium, whereas an unstable swing drives large, sustained fluctuations in voltage, current, and power and a widening rotor-angle separation that ends in loss of synchronism which ultimately requires controlled system separation \cite{rao2022synchrophasor}. Most asymmetrical swings remain stable, but those with an extended open-pole dead time can still evolve into unstable ones, so a protection scheme cannot assume symmetry and stability are correlated. Both failure modes have contributed to major cascading outages, and both worsen as a swing evolves from a stable, well-damped oscillation into an unstable one. A protection scheme must therefore perform two sequential discriminations: separating faults and faults occurring during a swing from swings alone, and, once a swing is confirmed, classifying it as stable or unstable so that blocking and out-of-step tripping logic can respond appropriately.

Conventional PSB schemes rely on the rate of change of impedance or resistance, using blinders and timers whose thresholds must be tuned per system and which struggle to generalize across operating conditions. This has motivated pattern-recognition alternatives, from wavelet- and Hilbert-transform-based swing detectors to machine-learning (ML) classifiers trained on positive/negative-sequence quantities, power-swing-center voltage, or FFT and statistical features of the 3-phase voltages and currents \cite{taheri2023overview}.


This remains an active research area. Taheri et al. \cite{taheri2023overview} catalogs many system and disturbance parameters that still affect detection reliability, and more recent work continues to refine both signal-processing and learning-based detectors. Sarajcev et al. \cite{sarajcev2024ml} proposed an SVM-based classifier to reinforce a generator's underimpedance and out-of-step relay logic , while ensemble methods -- XGBoost and Random Forest (RF) for transient-stability/instability classification \cite{senyuk2023transient}, and stacked CNN/ELM classifiers for two-stage transient stability assessment \cite{lei2024twostage} -- have shown consistent gains over single classifiers on related power-system stability tasks. At the same time, the increasing penetration of inverter-based resources is actively reshaping the problem: grid-forming and grid-following inverter control  changes swing and fault current signatures, motivating new detection strategies \cite{taheri2025hybrid, pallav2023}. Despite this most ML approaches to fault/swing discrimination rely on a limited and manually engineered feature set and a narrow classifier comparison evaluated by point-estimate accuracy without formal statistical testing of whether the compared classifiers actually differ \cite{amin2025}\cite{bera2021identification}\cite{sarajcev2024ml}.

Two developments in time-series ML can close this gap. First, automated feature-extraction libraries such as tsfresh \cite{christ2018tsfresh} and TSFEL \cite{barandas2020tsfel} compute hundreds of statistical, temporal, spectral, and fractal descriptors, paired with principled filtering (tsfresh's hypothesis-test-based selection, or general-purpose wrapper/embedded selectors). This combination has been applied to power-system diagnostics, e.g., tsfresh-based feature engineering for small-hydro condition monitoring \cite{desantis2023tsfresh} and SHAP-guided feature selection paired with a LightGBM classifier for power-transformer fault diagnosis \cite{liu2025shaplgbm}. Second, SHAP-based explainability \cite{lundberg2017shap} and rigorous multi-classifier statistical comparison (Friedman/Nemenyi omnibus-and-post-hoc procedure \cite{demsar2006} and the Nadeau--Bengio corrected resampled $t$-test \cite{nadeau2003}) are standard practice for benchmarking studies in ML, but have seen limited use in the power-swing/fault-discrimination literature.

This paper combines both developments and applies them to a 9-bus WSCC dataset used in \cite{bera2021identification}. To our knowledge, no prior fault/power-swing discrimination study has combined automated multi-library feature extraction with a statistically-grounded two-stage selection procedure and a formally validated multi-classifier benchmark; specifically, this paper contributes: (i) automated multi-library feature extraction, applying tsfresh, TSFEL, and tsfeatures to 3-phase relay voltages and currents; (ii) a two-stage feature-selection pipeline combining Boruta all-relevant filtering with RFECV and SHAP-guided minimal-optimal pruning; (iii) a benchmark of 14 classifiers spanning linear, kernel/instance-based, probabilistic, neural, and tree-ensemble/boosting families evaluated with repeated stratified cross-validation and validated by the Friedman/Nemenyi and Nadeau--Bengio statistical tests; (iv) application of the resulting pipeline to two cascaded classification stages, faults/faults-during-swing/power-swings and, subsequently, stable-versus-unstable swings; and (v) a SHAP-based interpretation of which signal characteristics drive the classification.

The remainder of the paper is organized as follows. Section~\ref{sec:simulation} describes the simulation and modeling. Section~\ref{sec:methodology} describes the feature extraction, selection, and classification methodology, and the classifiers benchmarked. Section~\ref{sec:results} presents the results for fault, fault-during-swing, \& power swing identification, and  classification of stable and unstable swings. Section~\ref{sec:conclusion} concludes the paper.


\begin{figure}[htbp]
\centerline{\includegraphics[width=3.51 in, height= 2.65 in]{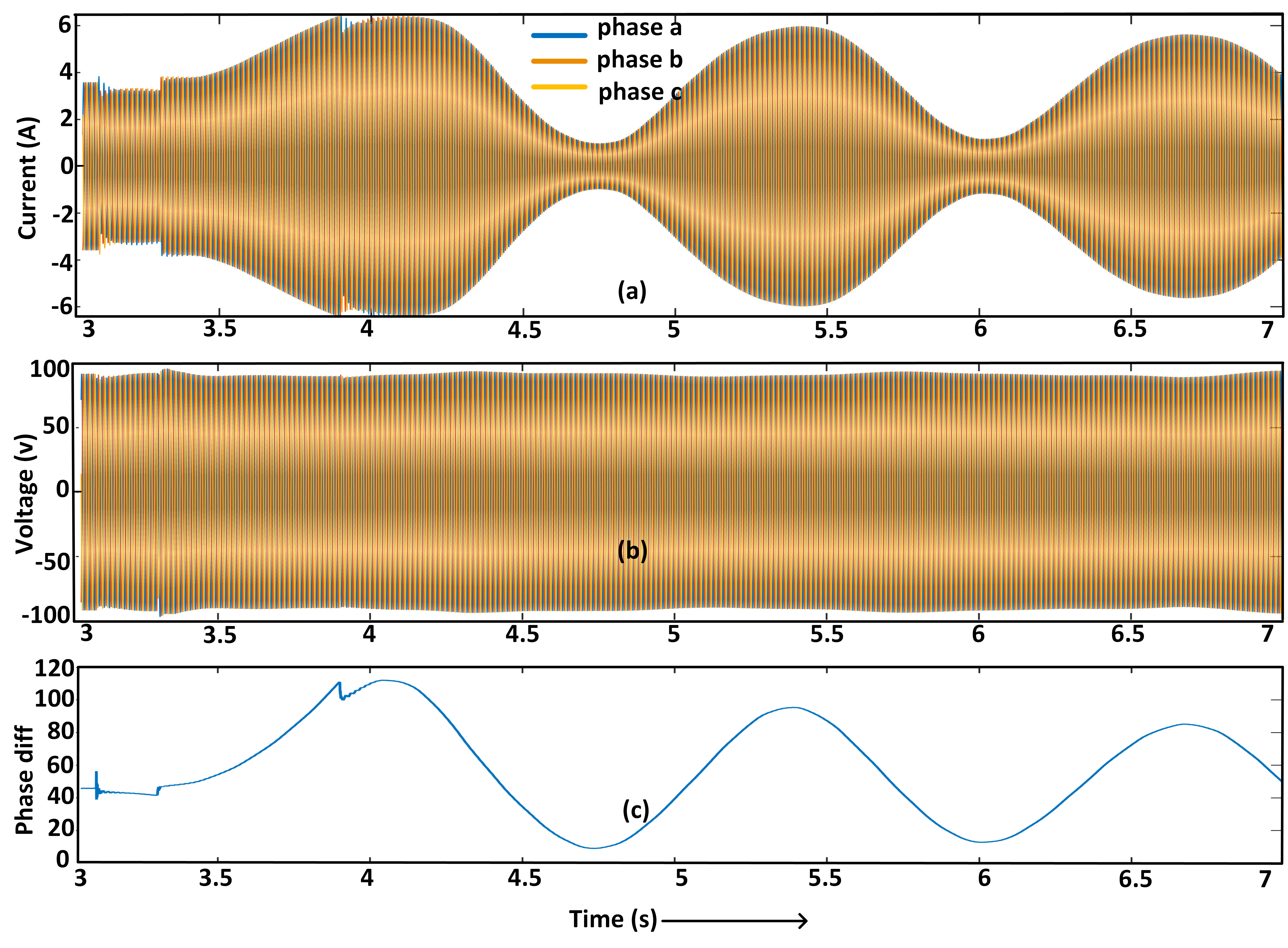}}
\caption{(a) 3-phase currents, (b) 3-phase voltages, (c) phase difference ($\phi$) for a typical symmetrical stable power swing \cite{bera2021identification}}
\label{ssps}
\end{figure}

\begin{figure}[htbp]
\centerline{\includegraphics[width=3.51 in, height= 2.65 in]{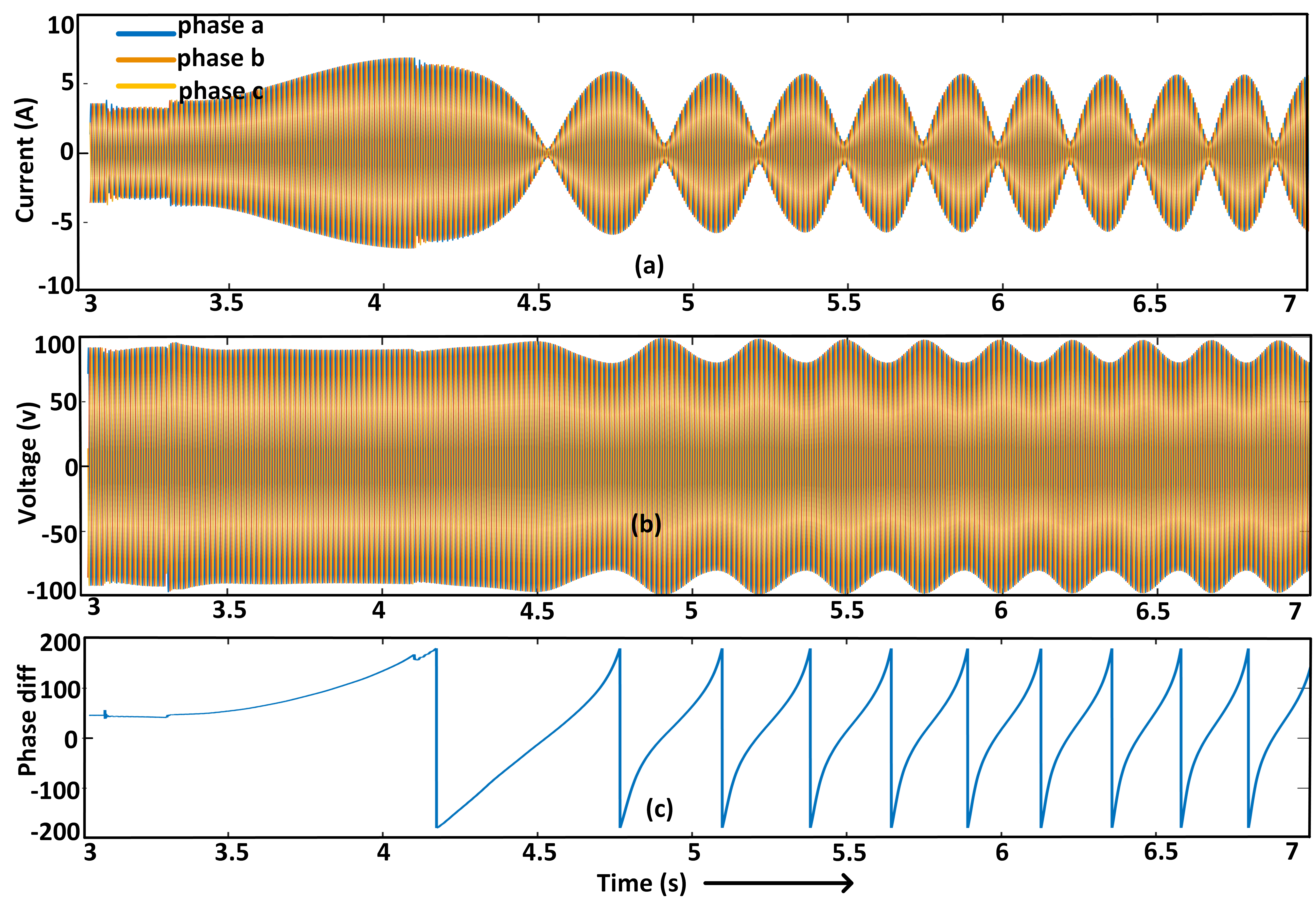}}
\caption{(a) 3-phase currents, (b) 3-phase voltages, (c) phase difference ($\phi$) for a typical symmetrical unstable power swing \cite{bera2021identification}}
\label{sups}
\end{figure}

\section{Simulation and Modeling}
\label{sec:simulation}

The cases used in this study were generated with the same power-system model and event set as the power-swing detection work \cite{bera2021identification}: a 9-bus, three-machine WSCC system (generation, transmission lines, transformers, and 3-phase loads operating at 60~Hz) modeled in PSCAD/EMTDC, with 3-phase voltage and current waveforms recorded at relay $R_{9-6}$ on bus~9. Three event categories relevant to the fault/power-swing discrimination problem addressed here were used: (i) transmission-line faults, obtained by varying fault type (single-, double-, and 3-phase), fault location, fault inception time, and fault resistance; (ii) faults occurring during an power swing, obtained by initiating a secondary fault partway through a simulated stable swing; and (iii) power swings alone, spanning symmetrical/asymmetrical and stable/unstable conditions, obtained by varying fault location, fault resistance, breaker clearing/reclosing times, and the phase-angle difference $\phi$ between bus~9 and bus~6. Table~\ref{tab:sim_cases} summarizes the resulting case counts. Detailed simulation parameters for each scenario are provided in \cite{bera2021identification}. One post-event cycle (84 samples at a 5~kHz sampling rate) of the 3-phase relay voltages and currents was used for feature extraction and classification. Fig.\ref{ssps} and \ref{sups} shows the phase-A, B, and C currents, the voltages seen by the relay, and $\phi$ between bus-9 and bus-6 for a symmetrical stable and an unstable power swings.

\begin{table}[htbp]
\centering
\caption{Simulated event cases used in this study.}
\label{tab:sim_cases}
\begin{tabular}{llc}
\toprule
Class & Sub-type & Cases \\
\midrule
Faults & T-line faults ($lg$, $llg$, $ll$, $lll$, $lllg$) & 1320 \\
\midrule
\multirow{2}{*}{Faults during swing} & Symmetrical stable swing & 660 \\
                                     & Asymmetrical stable swing & 660 \\
\midrule
\multirow{4}{*}{Power swings} & Symmetrical, stable & 720 \\
                              & Symmetrical, unstable & 540 \\
                              & Asymmetrical, stable & 450 \\
                              & Asymmetrical, unstable & 495 \\
\midrule
\textbf{Total} & & \textbf{4845} \\
\bottomrule
\end{tabular}
\end{table}

\section{Methodology}
\label{sec:methodology}

\subsection{Automated Multi-Library Feature Extraction}
\label{sec:methodology_extraction}

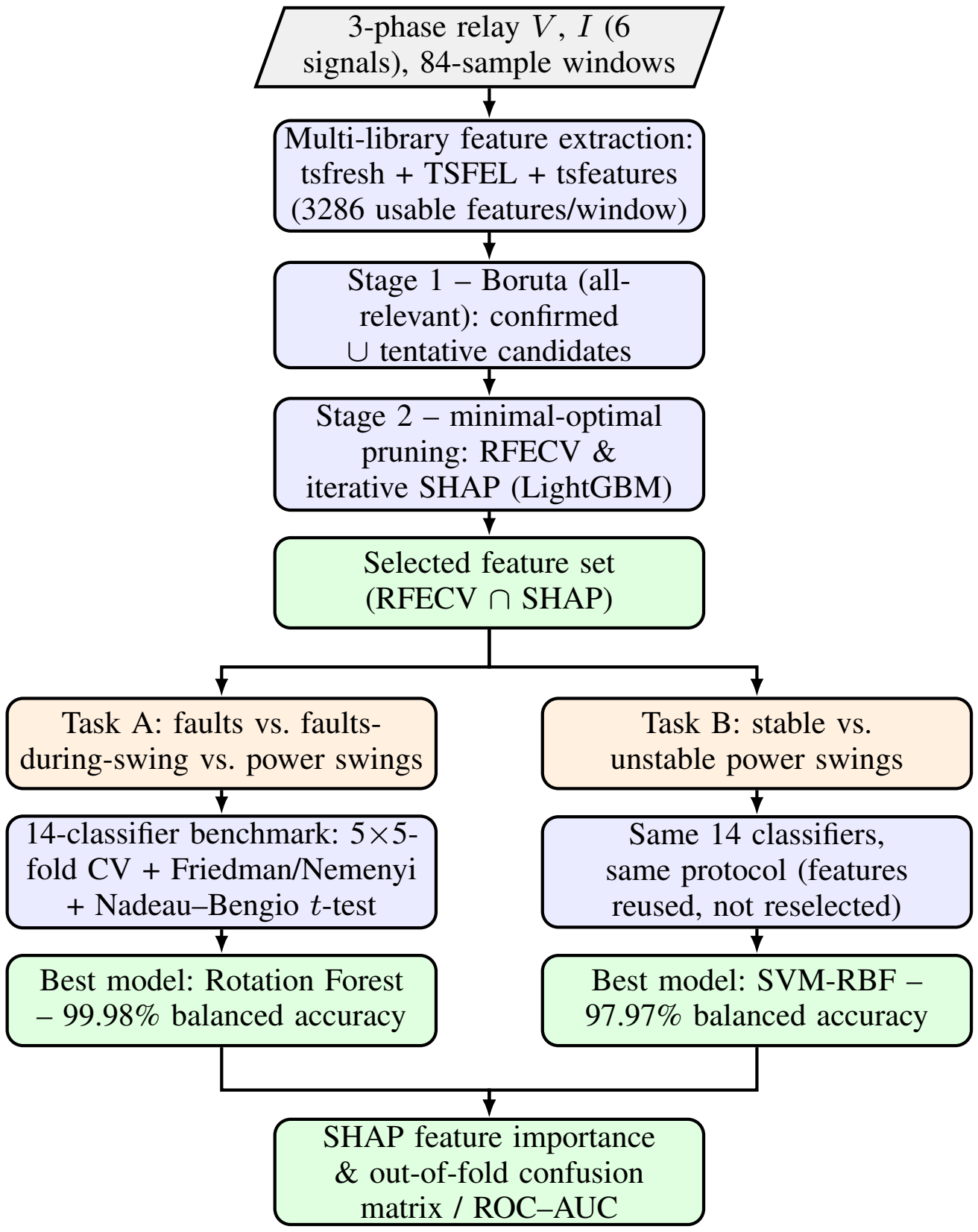
\begin{figure}[htbp]
\centering
\begin{tikzpicture}[ scale=0.95,
    process/.style = {rectangle, rounded corners, draw=black, thick,
                       fill=blue!8, minimum width=38 mm, minimum height=8mm,
                       align=center, font=\footnotesize, text width=37mm,
                       inner sep=1.5pt},
    io/.style      = {trapezium, trapezium left angle=70, trapezium right angle=110,
                       draw=black, thick, fill=gray!12,
                       minimum width=39mm, minimum height=7mm,
                       align=center, font=\footnotesize, text width=34mm,
                       inner sep=1.5pt},
    decision/.style= {rectangle, rounded corners, draw=black, thick,
                       fill=orange!12, minimum width=38mm, minimum height=8mm,
                       align=center, font=\footnotesize, text width=37mm,
                       inner sep=1.5pt},
    result/.style  = {rectangle, rounded corners, draw=black, thick,
                       fill=green!12, minimum width=36mm, minimum height=8mm,
                       align=center, font=\footnotesize, text width=37mm,
                       inner sep=1.5pt},
    arr/.style     = {-{Latex[length=1.6mm]}, thick},
]

\node[io]      (sim)     at (0, 0)     { 3-phase relay $V$, $I$ (6 signals), 84-sample windows};
\node[process] (extract) at (0,-1.20) {Multi-library feature extraction: tsfresh + TSFEL + tsfeatures (3286 usable features/window)};
\node[process] (boruta)  at (0,-2.5) {Stage 1 -- Boruta (all-relevant): confirmed $\cup$ tentative candidates};
\node[process] (wrapper) at (0,-3.8) {Stage 2 -- minimal-optimal pruning: RFECV \& iterative SHAP (LightGBM)};
\node[result]  (selected) at (0,-5.00) {Selected feature set (RFECV $\cap$ SHAP)};

\draw[arr] (sim)     -- (extract);
\draw[arr] (extract) -- (boruta);
\draw[arr] (boruta)  -- (wrapper);
\draw[arr] (wrapper) -- (selected);

\node[decision] (taskA) at (-2.5,-6.5) {Task A: faults vs.\ faults-during-swing vs.\ power swings};
\node[decision] (taskB) at ( 2.5,-6.5) {Task B: stable vs.\ unstable power swings};

\node[process]  (clfA) at (-2.5,-7.7) {14-classifier benchmark: 5$\times$5-fold CV + Friedman/Nemenyi + Nadeau--Bengio $t$-test};
\node[process]  (clfB) at (2.5,-7.7) {Same 14 classifiers, same protocol (features reused, not reselected)};

\node[result]   (resA) at (-2.5,-8.9) {Best model: Rotation Forest -- 99.98\% balanced accuracy};
\node[result]   (resB) at (2.5,-8.9) {Best model: SVM-RBF -- 97.97\% balanced accuracy};

\draw[arr] (selected.south) -- ++(0,-0.35) -| (taskA.north);
\draw[arr] (selected.south) -- ++(0,-0.35) -| (taskB.north);
\draw[arr] (taskA) -- (clfA);
\draw[arr] (taskB) -- (clfB);
\draw[arr] (clfA)  -- (resA);
\draw[arr] (clfB)  -- (resB);

\node[result] (interp) at (0,-10.5) {SHAP feature importance \& out-of-fold confusion matrix / ROC--AUC};
\draw[arr] (resA.south) -- ++(0,-0.4) -| (interp.north);
\draw[arr] (resB.south) -- ++(0,-0.4) -| (interp.north);

\end{tikzpicture}
\caption{Overview of the proposed pipeline}
\label{fig:pipeline}
\end{figure}

Each of the six 3-phase relay signals (currents $C_x$, $C_y$, $C_z$ and voltages $V_x$, $V_y$, $V_z$) was pre-segmented into 4845 fixed-length, non-overlapping windows of 84 samples, matching one relay processing cycle. Three complementary automated feature-extraction libraries were applied independently to every channel and window: tsfresh~\cite{christ2018tsfresh} under its efficient configuration (on the order of 750 statistical, temporal, and spectral characteristics per channel), TSFEL~\cite{barandas2020tsfel} restricted to its statistical and temporal feature domains, and tsfeatures, a list of roughly 38 forecasting-oriented meta-features per channel (e.g., Hurst exponent, autocorrelation- and entropy-based measures), computed with seasonal-decomposition-dependent features disabled since a single 84-sample window carries no known periodicity. Tsfresh has been used to extract relevant features from power system faults previously in \cite{pedespallav}\cite{hifpallav2024}. All three libraries were run across all six channels and concatenated into one per-window feature table. Columns that were constant, undefined for every window, or non-finite were dropped, and remaining missing values were median-imputed, leaving 3286 usable features per window. The overall pipeline covering both the primary task (\textit{Task A}: Fault, fault-during-swing, \& power swing identification) and the secondary task (\textit{Task B}: Stable \& unstable power swing classification) is summarized in Fig.~\ref{fig:pipeline}.

\subsection{Two-Stage Feature Selection}
\label{sec:methodology_selection}

The 3286-feature table was reduced with a two-stage procedure combining an all-relevant filter with a minimal-optimal wrapper, since the two serve different purposes: the former retains every feature carrying genuine signal (useful for interpretation but too redundant for deployment), while the latter alone can be unstable when starting from a feature set this wide. 

Stage 1 applied Boruta~\cite{kursa2010boruta}, an all-relevant selector that wraps a RF (class-balanced, max depth 6) and iteratively compares each feature's importance against that of a shuffled ("shadow") copy of itself, retaining every feature that consistently outperforms its shadow. Features confirmed or left tentative by Boruta were carried forward as the candidate set. Stage 2 then pruned this candidate set to a minimal-optimal subset by two independent wrapper methods built around a gradient-boosted model (LightGBM~\cite{ke2017lightgbm}): (i) recursive feature elimination with cross-validation (RFECV~\cite{guyon2002rfe}), removing the weakest $\sim$5\% of remaining features per step and selecting the subset size that maximizes cross-validated balanced accuracy; and (ii) an iterative SHAP-pruning procedure that repeatedly fits LightGBM, ranks features by mean absolute SHAP value, discards the weakest quartile, and re-evaluates cross-validated balanced accuracy at each round, ultimately keeping the smallest subset whose score falls within one standard error of the best round observed. Both stratified cross-validation splits used the same fold structure ($k$ adapting to the smallest class). The RFECV set, the SHAP set, and their intersection were all retained; the intersection -- the most conservative of the three -- was used as the default feature set for downstream classification. 

\subsection{Classifier Benchmark and Statistical Comparison}
\label{sec:methodology_classification}

The feature set retained after the Boruta/RFECV--SHAP selection stage was evaluated with a benchmark of 14 tabular classifiers.
The classifiers (Table~\ref{tab:classifiers}) were grouped into five families by their underlying decision mechanism. \textit{Linear} models (Logistic Regression (LR), Ridge, Linear SVM (LSVM)) separate classes with a hyperplane in the original feature space and serve as a baseline for how linearly separable the selected features are. \textit{Tree ensembles} combine many decision trees, either through bagging/randomization (RF, Extra Trees (ET), ROF) or sequential boosting (Gradient Boosting (GB), LightGBM, XGBoost, CatBoost); this group is the largest because it also covers the classifiers used internally by state-of-the-art time-series-classification pipelines (ROF in FreshPRINCE, ET ensembles in HIVE-COTE~2). \textit{Kernel/instance-based} models (SVM-RBF, $k$NN) classify based on similarity or distance to training samples rather than an explicit decision boundary. \textit{Probabilistic} (Naive Bayes (NB)) and \textit{neural} (MLP) form single-classifier groups representing a generative baseline and a nonlinear connectionist model, respectively.

All models were evaluated under an identical repeated stratified $k$-fold cross-validation protocol, with $k=5$ folds repeated 5 times (25 total fold evaluations), scored on balanced accuracy and macro-averaged $F_1$ to account for the residual class imbalance between fault types. Model comparison followed a two-stage statistical procedure: an omnibus Friedman test on the per-fold balanced-accuracy scores of all classifiers, followed by a Nemenyi post-hoc test to identify which pairs of classifiers differ significantly \cite{demsar2006}; in addition, the top-ranked classifier was compared against every other classifier with a Nadeau--Bengio corrected resampled paired $t$-test \cite{nadeau2003}, which inflates the variance estimate to account for the non-independence of cross-validation folds, with $p$-values adjusted for multiple comparisons using the Holm--Bonferroni step-down procedure. The best-performing classifier was refit on the full dataset for deployment, while all reported confusion-matrix, per-class, and ROC--AUC statistics were computed from out-of-fold predictions obtained via cross-validated prediction rather than from the refit model, avoiding optimistic bias from evaluating on training data. Feature importance for the selected model was obtained from its native importance scores where available, or otherwise from permutation-based SHAP values (mean absolute SHAP value across a representative subsample).

\subsection{Stable/Unstable Swing Classification via Feature Reuse}
\label{sec:methodology_stable_unstable}

The \textit{Task B} reused the same feature set selected for (\textit{Task A}) (Section~\ref{sec:methodology_selection}) rather than re-running feature selection. Reusing the same features isolates the effect of the label (three-way fault type versus binary swing stability) from the effect of feature choice, keeping the two studies directly comparable. 

The stable/unstable feature table was extracted following the same procedure (Section~\ref{sec:methodology_extraction}) and then restricted to the features retained by the intersection selection. The classifier benchmark, cross-validation protocol, and statistical comparison procedure of Section~\ref{sec:methodology_classification} were then applied unchanged to this binary target.



\begin{table}[htbp]
\centering
\caption{Candidate Classifiers grouped by model family.}
\label{tab:classifiers}
\begin{tabular}{lll}
\toprule
Family & Classifier & Implementation \\
\midrule
\multirow{3}{*}{Linear} & LR & scikit-learn \\
                        & Ridge & scikit-learn \\
                        & LSVM & scikit-learn \\
\midrule
\multirow{7}{*}{Tree ensemble} & RF & scikit-learn \\
                               & ET & scikit-learn \\
                               & ROF & aeon \\
                               & GB & scikit-learn \\
                               & LightGBM & LightGBM \\
                               & XGBoost & XGBoost \\
                               & CatBoost & CatBoost \\
\midrule
\multirow{2}{*}{Kernel / instance-based} & SVM (RBF kernel) & scikit-learn \\
                                        & $k$-Nearest Neighbor & scikit-learn \\
\midrule
Probabilistic & NB (Gaussian) & scikit-learn \\
\midrule
Neural & MLP & scikit-learn \\
\bottomrule
\end{tabular}
\end{table}

\section{Results}
\label{sec:results}
\subsection{Faults, Faults during Power-Swing, \& Power Swing Identification (\textbf{Task A})}
\label{sec:results_faults_ps}
Table~\ref{tab:leaderboard} summarises the cross-validated performance of all 14 classifiers for \textit{Task A:} Faults, faults-during-swing, \& power swing identification, ranked according to mean balanced accuracy. ROF achieved the highest balanced accuracy ($99.98\% \pm 0.04\%$, 95\% CI $[99.96\%, 100.00\%]$), followed closely by ET ($99.97\%$), RF ($99.97\%$), and the gradient-boosting models, all differing by less than $0.10$ percentage points. In contrast, the linear and probabilistic classifiers (LR, LSVM, Ridge, and NB) produced substantially lower accuracies, indicating that the fault classes are not linearly separable within the selected feature space.

\begin{table}[htbp]
\centering
\caption{Cross-validated classifier leaderboard for \textit{Task A}. 95\% confidence interval on balanced accuracy from the $t$-distribution over CV folds.}
\label{tab:leaderboard}
\setlength{\tabcolsep}{3pt}
\renewcommand{\arraystretch}{0.95}
\begin{tabular}{lcccc}
\toprule
Classifier & Bal.\ Acc.\ (\%) & Macro-$F_1$ (\%) & 95\% CI (\%) & Fit time (s) \\
\midrule
\textbf{ROF}      & $99.98 \pm 0.04$ & $99.97 \pm 0.06$ & [99.96, 100.00] & 12.73 \\
ET           & $99.97 \pm 0.04$ & $99.96 \pm 0.06$ & [99.95, 99.99] & 1.35 \\
RF         & $99.97 \pm 0.04$ & $99.96 \pm 0.06$ & [99.95, 99.99] & 11.74 \\
XGBoost              & $99.95 \pm 0.06$ & $99.94 \pm 0.07$ & [99.92, 99.97] & 0.97 \\
GB     & $99.94 \pm 0.09$ & $99.93 \pm 0.10$ & [99.90, 99.98] & 21.63 \\
CatBoost             & $99.93 \pm 0.07$ & $99.92 \pm 0.09$ & [99.91, 99.96] & 5.10 \\
LightGBM             & $99.92 \pm 0.06$ & $99.91 \pm 0.07$ & [99.90, 99.95] & 9.22 \\
SVM-RBF              & $99.92 \pm 0.07$ & $99.91 \pm 0.08$ & [99.89, 99.95] & 0.24 \\
MLP                  & $99.87 \pm 0.16$ & $99.86 \pm 0.16$ & [99.80, 99.93] & 1.58 \\
$k$NN                & $99.73 \pm 0.19$ & $99.71 \pm 0.21$ & [99.64, 99.81] & 0.00 \\
LR  & $96.46 \pm 0.51$ & $96.39 \pm 0.52$ & [96.25, 96.67] & 0.10 \\
LSVM           & $94.88 \pm 0.57$ & $94.83 \pm 0.58$ & [94.64, 95.12] & 0.17 \\
Ridge                & $86.14 \pm 0.98$ & $86.32 \pm 1.04$ & [85.73, 86.55] & 0.01 \\
NB          & $80.84 \pm 1.23$ & $79.88 \pm 1.37$ & [80.32, 81.36] & 0.00 \\
\bottomrule
\end{tabular}
\end{table}

The Friedman test rejected the null hypothesis that all classifiers perform equally ($\chi^2 = 262.26$, $p = 1.81\times10^{-48}$, $n=25$ fold evaluations), confirming statistically significant differences among the 14 classifiers. Both the Nemenyi post-hoc test and the Nadeau--Bengio corrected paired $t$-tests identified Ridge, NB, LSVM, and LR as performing significantly worse than ROF (Holm-adjusted $p < 10^{-10}$). In contrast, ET, RF, XGBoost, GradientBoosting, CatBoost, LightGBM, SVM-RBF, and MLP were statistically indistinguishable from the best-performing model (Holm-adjusted $p > 0.95$). Although $k$NN showed a nominally significant difference before multiple-comparison correction ($p=0.02$), it was no longer significant after Holm adjustment ($p=0.16$). The corrected pairwise comparisons are summarised in Table~\ref{tab:ttest}.

\begin{table}[htbp]
\centering
\caption{Nadeau--Bengio corrected paired $t$-test comparing \textbf{ROF} with each competing classifier (25 fold evaluations, $k=5$).}
\label{tab:ttest}
\begin{tabular}{lccc}
\toprule
Classifier & $\Delta$ Bal.\ Acc.\ (pp) & $t$ & $p_\text{Holm}$ \\
\midrule
NB          & 19.14 & 28.22 & $<10^{-3}$ \\
Ridge                & 13.84 & 25.92 & $<10^{-3}$ \\
LSVM           & 5.10  & 16.45 & $1.56\times10^{-13}$ \\
LR  & 3.52  & 12.81 & $3.18\times10^{-11}$ \\
$k$NN                & 0.25  & 2.55  & 0.16 \\
LightGBM             & 0.05  & 1.61  & 0.96 \\
MLP                  & 0.11  & 1.37  & 1.00 \\
SVM-RBF              & 0.06  & 1.30  & 1.00 \\
CatBoost             & 0.04  & 1.09  & 1.00 \\
XGBoost              & 0.03  & 0.81  & 1.00 \\
GB    & 0.04  & 0.75  & 1.00 \\
RF         & 0.01  & 0.61  & 1.00 \\
ET           & 0.01  & 0.47  & 1.00 \\
\bottomrule
\end{tabular}
\end{table}

Out-of-fold predictions obtained using ROF achieved an overall classification accuracy of $99.98\%$ (Table~\ref{tab:classreport}). Fault class 1 was classified perfectly, while classes 2 and 3 exhibited near-perfect precision and recall. The corresponding one-vs-rest ROC analysis further demonstrated excellent class separability, with every class and the macro average achieving an AUC of $1.00$ (Table~\ref{tab:roc}).

\begin{table}[htbp]
\centering
\caption{Out-of-fold classification report for \textbf{ROF} (\textit{Task A}).}
\label{tab:classreport}
\begin{tabular}{lcccc}
\toprule
Class & Precision & Recall & $F_1$ & Support \\
\midrule
1 & 1.00 & 1.00 & 1.00 & 1320 \\
2 & 1.00 & 1.00 & 1.00 & 1320 \\
3 & 1.00 & 1.00 & 1.00 & 2205 \\
\midrule
Macro avg. & 1.00 & 1.00 & 1.00 & 4845 \\
Accuracy & \multicolumn{4}{c}{1.00} \\
\bottomrule
\end{tabular}
\end{table}

\begin{table}[htbp]
\centering
\caption{One-vs-rest ROC-AUC for \textbf{ROF} model.}
\label{tab:roc}
\begin{tabular}{lc}
\toprule
Class & AUC \\
\midrule
1 & 1.00 \\
2 & 1.00 \\
3 & 1.00 \\
\midrule
Macro avg. & 1.00 \\
\bottomrule
\end{tabular}
\end{table}

Table~\ref{tab:importance} presents the 18 most important features used as input ranked by mean absolute SHAP value. The dominant predictors are trend- and autocorrelation-based descriptors extracted primarily from the $V_y$, $C_x$, and $C_z$ channels. These results indicate that fault-induced transients are characterised predominantly by changes in local temporal trends and short-lag autocorrelation rather than by spectral information alone.

\begin{table}[htbp]
\centering
\caption{18 selected features by mean $|\mathrm{SHAP}|$ for the selected model.}
\label{tab:importance}
\setlength{\tabcolsep}{2pt}
\renewcommand{\arraystretch}{0.95}
\begin{tabular}{cllc}
\toprule
 Signal & Feature & Library & Importance \\
\midrule
 $V_y$ & Autocorrelation (lag 10) of first difference & tsfeatures & 0.05 \\
 $V_y$ & Max.\ $R^2$ of linear trend (5-sample chunk) & tsfresh & 0.04 \\
 $C_z$ & Variance of trend intercept(50-sample chunk) & tsfresh & 0.04 \\
 $C_x$ & Variance of trend std.\ error(5-sample chunk) & tsfresh & 0.04 \\
 $V_y$ & Mean $R^2$ of linear trend (10-sample chunk) & tsfresh & 0.04 \\
 $C_x$ & AR(10) coefficient 1 & tsfresh & 0.03 \\
 $C_z$ & Slope of linear trend (full window) & tsfresh & 0.03 \\
 $C_z$ & Mean slope of linear trend(50-sample chunk) & tsfresh & 0.03 \\
 $C_x$ & CWT coefficient 14 (width 20) & tsfresh & 0.03 \\
 $V_y$ & $|$FFT coefficient$|$ at index 4 & tsfresh & 0.03 \\
 $V_x$ & Stability (variance of tiled-window means) & tsfeatures & 0.03\\
 $V_z$ & $|$FFT coefficient$|$ at index 2 & tsfresh & 0.03 \\
 $C_y$ & CWT coefficient 3 (width 10) & tsfresh & 0.02 \\
 $C_x$ & Max.\ trend intercept (50-sample chunk) & tsfresh & 0.02 \\
 $V_y$ & AR(10) coefficient 1 & tsfresh & 0.02\\
 $C_z$ & AR(10) coefficient 2 & tsfresh & 0.02 \\
 $V_x$ & Mean abs.\ change, 40-60th percentile band & tsfresh & 0.02 \\
 $C_y$ & Spectral centroid of FFT amplitude & tsfresh & 0.01 \\
\bottomrule
\end{tabular}
\end{table}

\subsection{Stable \& Unstable Power Swing Classification (\textbf{Task B})}
\label{sec:results_stable_unstable}

Using the same 14 classifiers, feature set, and repeated $5\times5$-fold stratified cross-validation protocol as \textit{Task A}, Table~\ref{tab:leaderboard_su} summarises performance on \textit{Task B:} Stable \& unstable power-swing classification. SVM-RBF ranked first, with a balanced accuracy of $97.97\% \pm 0.80\%$ (95\% CI $[97.63\%, 98.31\%]$), closely followed by RF, LSVM, and CatBoost.

\begin{table}[htbp]
\centering
\caption{Cross-validated classifier leaderboard for \textit{Task B}.}
\label{tab:leaderboard_su}
\setlength{\tabcolsep}{3pt}
\renewcommand{\arraystretch}{1}
\begin{tabular}{lccccc}
\toprule
Classifier & Bal.\ Acc.\ (\%) & Macro-$F_1$ (\%) & 95\% CI (\%) & Fit time (s) \\
\midrule
\textbf{SVM-RBF}               & $97.97 \pm 0.80$ & $97.86 \pm 0.85$ & [97.63, 98.31] & 0.09 \\
RF          & $97.90 \pm 0.79$ & $97.83 \pm 0.82$ & [97.57, 98.23] & 4.12 \\
LSVM            & $97.76 \pm 0.86$ & $97.67 \pm 0.90$ & [97.39, 98.12] & 0.01 \\
CatBoost              & $97.73 \pm 0.94$ & $97.67 \pm 0.95$ & [97.33, 98.12] & 2.88 \\
LightGBM              & $97.62 \pm 0.97$ & $97.58 \pm 0.99$ & [97.21, 98.03] & 4.79 \\
XGBoost               & $97.60 \pm 1.21$ & $97.56 \pm 1.21$ & [97.10, 98.11] & 0.25 \\
LR   & $97.59 \pm 0.82$ & $97.54 \pm 0.84$ & [97.25, 97.94] & 0.01 \\
ET            & $97.52 \pm 0.79$ & $97.47 \pm 0.83$ & [97.19, 97.86] & 0.81 \\
ROF       & $97.52 \pm 0.87$ & $97.47 \pm 0.90$ & [97.16, 97.89] & 4.18 \\
GB      & $97.39 \pm 1.00$ & $97.33 \pm 1.01$ & [96.97, 97.81] & 3.12 \\
MLP                   & $97.22 \pm 0.98$ & $97.18 \pm 0.98$ & [96.81, 97.64] & 1.28 \\
$k$NN                 & $97.21 \pm 0.87$ & $97.16 \pm 0.88$ & [96.84, 97.57] & 0.00 \\
Ridge                 & $96.64 \pm 0.81$ & $96.61 \pm 0.82$ & [96.29, 96.98] & 0.00 \\
NB           & $88.86 \pm 1.33$ & $88.14 \pm 1.43$ & [88.30, 89.42] & 0.00 \\
\bottomrule
\end{tabular}
\end{table}

The Friedman test rejected equal performance across all 14 classifiers ($\chi^2 = 151.88$, $p = 8.63\times10^{-26}$, $n=25$ fold evaluations), confirming that at least one classifier differs meaningfully from the rest despite the tight overall spread in Table~\ref{tab:leaderboard_su}. The Nadeau--Bengio corrected paired 
$t$-test (Table~\ref{tab:ttest_su}) identifies only NB and Ridge as significantly inferior to SVM-RBF (Holm-adjusted $p<0.05$). The remaining 11 classifiers (tree-ensemble, boosting, kernel, and neural method tested) are statistically indistinguishable from SVM-RBF on this task. The rank-based Nemenyi post-hoc test tells a partly different story: it additionally flags ET, GB, $k$NN, MLP, and ROF as significantly different from SVM-RBF. This divergence is expected -- Nemenyi compares average ranks across all 14 classifiers simultaneously and can be sensitive to small but consistent rank differences even when the corresponding score gap is small, whereas the Nadeau--Bengio test directly compares score magnitudes with a variance correction for non-independent folds. 

\begin{table}[htbp]
\centering
\caption{Nadeau--Bengio corrected paired $t$-test, \textbf{SVM-RBF} vs.\ each other classifier, Holm-adjusted (25 fold evaluations, $k=5$).}
\label{tab:ttest_su}
\begin{tabular}{lccc}
\toprule
Classifier & $\Delta$ Bal.\ Acc.\ (pp) & $t$ & $p_\text{Holm}$ \\
\midrule
NB          & 9.11 & 13.08 & $< 0.001$ \\
Ridge                & 1.33 & 3.37  & 0.030 \\
$k$NN                & 0.76 & 2.32  & 0.324 \\
MLP                  & 0.75 & 2.08  & 0.485 \\
GB     & 0.58 & 1.76  & 0.828 \\
ET          & 0.45 & 1.73  & 0.828 \\
ROF     & 0.45 & 1.57  & 0.913 \\
LR  & 0.38 & 1.36  & 1.000 \\
LSVM           & 0.21 & 1.20  & 1.000 \\
LightGBM             & 0.35 & 0.96  & 1.000 \\
XGBoost              & 0.37 & 0.77  & 1.000 \\
CatBoost             & 0.25 & 0.71  & 1.000 \\
RF        & 0.07 & 0.34  & 1.000 \\
\bottomrule
\end{tabular}
\end{table}

\begin{figure}[htbp]
\centerline{\includegraphics[width=2.9 in, height= 1.7 in]{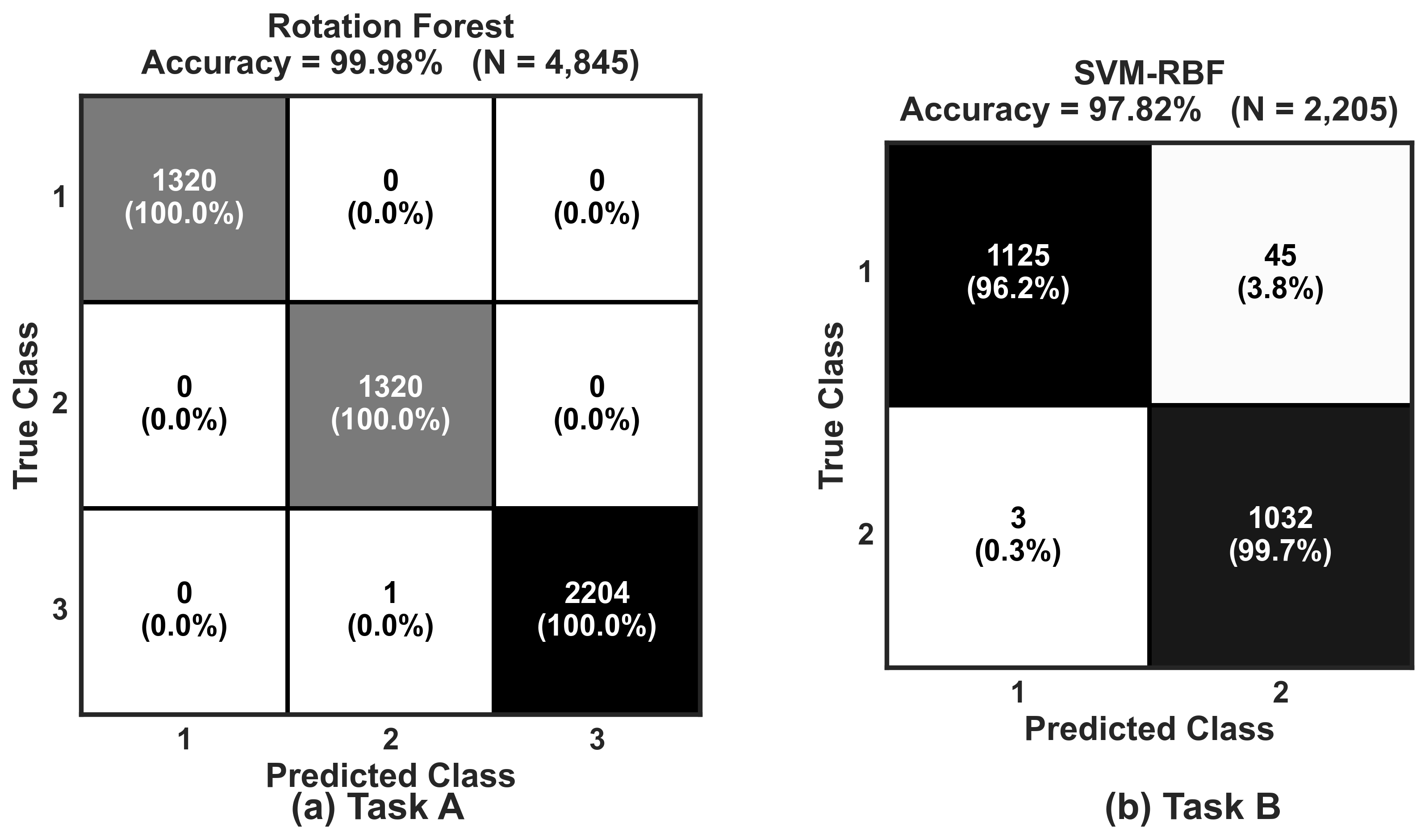}}
\caption{(a) Confusion matrix Task A, (b) Confusion matrix Task B,}
\label{conf}
\end{figure}

Out-of-fold predictions from the selected model (SVM-RBF) yield an overall accuracy of $97.82\%$ (Table~\ref{tab:classreport_su}); precision and recall are asymmetric across classes, with class 1 recalled less often ($96.15\%$) but predicted more precisely ($99.73\%$) than class 2. The one-vs-rest ROC analysis nonetheless shows strong separability, with an AUC of $99.34\%$ for both classes (Table~\ref{tab:roc_su}). Fig. \ref{conf} shows the confusion matrix for \textit{Task A} and \textit{Task B}.

\begin{table}[h]
\centering
\caption{Out-of-fold classification report for \textbf{SVM-RBF} (\textit{Task B})}
\label{tab:classreport_su}
\begin{tabular}{lcccc}
\toprule
Class & Precision (\%) & Recall (\%) & $F_1$ (\%) & Support \\
\midrule
1 & 99.73 & 96.15 & 97.91 & 1170 \\
2 & 95.82 & 99.71 & 97.73 & 1035 \\
\midrule
Macro avg. & 97.78 & 97.93 & 97.82 & 2205 \\
Accuracy   & \multicolumn{4}{c}{97.82} \\
\bottomrule
\end{tabular}
\end{table}

\begin{table}[ht]
\centering
\caption{One-vs-rest ROC-AUC for \textbf{SVM-RBF} model.}
\label{tab:roc_su}
\begin{tabular}{lc}
\toprule
Class & AUC (\%) \\
\midrule
1 & 99.34 \\
2 & 99.34 \\
\midrule
Macro avg. & 99.34 \\
\bottomrule
\end{tabular}
\end{table}



\section{Conclusion}
\label{sec:conclusion}

This paper presents an automated pipeline: multi-library feature extraction (tsfresh, TSFEL, tsfeatures) across all six relay voltage/current channels, a two-stage Boruta/RFECV/SHAP selection, and a 14-classifier benchmark validated by formal statistical testing.
On the primary \textit{Task A} (identifying faults, faults during a power swing, and power swings across 4845 cases with 18 features) ROF led at 99.98\% balanced accuracy, with a significant Friedman test ($p\approx2\times10^{-48}$) and 9 of the other 13 classifiers statistically indistinguishable from it. Using the same 18 features for the subsequent stable-versus-unstable swing \textit{Task B}, SVM-RBF led at 97.97\% balanced accuracy (Friedman $\chi^2=151.88$, $p=8.63\times10^{-26}$); here  NB and Ridge were significantly worse.
SHAP-based importance on the top models highlighted trend- and autocorrelation-based descriptors of specific voltage/current channels as the dominant features. In practice, the statistical ties among most ensemble, boosting, and kernel models are useful for deployment: several tied alternatives train substantially faster than the top-ranked model, making them better suited to real-time use without any significant loss in accuracy.
The primary limitation of this study is that the proposed framework was evaluated only on IEEE 9-bus test system under ideal operating conditions. Future research should investigate its robustness under measurement noise, diverse operating scenarios, and inverter-based-resource-dominated networks, while extending the validation to larger systems, such as the IEEE 39-bus.

\balance
\bibliographystyle{IEEEtran}
\bibliography{ref}
\end{document}